\documentclass[12pt,onecolumn]{article}
\usepackage{graphics}
\usepackage{amssymb,epsfig,amsmath,euscript,array,cite}
 \usepackage{setspace}  

\newcommand{\head}[1]{\subsection {  #1}}

\newcounter{multieqs}

\newcommand{\bq}{\begin{equation}}
\newcommand{\fq}{\end{equation}}
\newcommand{\bqr}{\begin{eqnarray}}
\newcommand{\fqr}{\end{eqnarray}}

\newcommand{\be}{\begin{equation}}
\newcommand{\ee}{\end{equation}}
\newcommand{\eq}[1]{(\ref{#1})}

\newcommand{\bra}[1]{\langle #1|}
\newcommand{\ket}[1]{|#1 \rangle}

\newcommand{\bm}[1]{\mbox{\boldmath $#1$}}

\def\nn{\nonumber}
\def\bea{\begin{eqnarray}}
\def\eea{\end{eqnarray}}
\let\bm=\bibitem

\def\beqa{\begin{eqnarray}} 
\def\eeqa{\end{eqnarray}} 
\def\beq{\begin{equation}} 
\def\eeq{\end{equation}}

\def\one{\mbox{1 \kern-.59em {\rm l}}}

\def\a{\alpha}		
\def\b{\beta}		

\def\d{\delta}		
\def\e{\epsilon}

\def\l{\lambda}	\def\L{\Lambda}
\def\m{\mu}	\def\n{\nu}

\def\s{\sigma}	\def\S{\Sigma}
\def\t{\tau}
\def\th{\theta}

 \def\cN{{\cal N}}

\def\bq{{\bf q}}

\def\d{\delta}

 \def\del{\partial}

\def\one{1\!\!1\,\,}

\newcommand{\tr}{\mbox{tr}}

\def\ZZ{\mathbb{Z}}
\def\RR{\mathbb{R}}

\def\bcomment#1{}

 \def\ii{{\rm i}}

\def\uno{\mbox{1 \kern-.59em {\rm l}}}

\begin{document}

\hfill {\tt hep-th/0502167}

\begin{center}
 
{\large \bf NON-COMMUTATIVE GEOMETRY FROM STRINGS}
\footnote{To appear in
{\sl Encyclopedia of Mathematical Physics}, J.-P.~Fran\c{c}oise, 
G.~Naber and T.S.~Tsou, eds., Elsevier, 2006.}

\vspace{20pt}
 
{\large Chong-Sun Chu}
 
{\em Department of Mathematical Sciences, 
University of Durham,  DH1 3LE, UK}

\sffamily{chong-sun.chu@durham.ac.uk}
 

\end{center}
  


\noindent {\large \bf Synopsis}

One of 
the most important recent developments in string theory is the discovery
of D-branes and a deeper understanding of their properties. Among other
things, this allows one to derive for the first time noncommutative
geometry from  string theory.
The resulting noncommutative geometry has stimulated a wide range of
physical applications and has leaded to new insights into a fundamental
formulation of string theory.


\section{Noncommutative Geometry from String Theory}

The first use of noncommutative geometry in string theory appears in the
work of Witten on open string field theory \cite{witten1} where the
noncommutativity is associated with the product of open string fields.
Noncommutative geometry appears in the recent development of string
theory in the seminal work of Connes, Douglas and Schwarz \cite{cds}
where they constructed and identified the compactification of Matrix
theory on noncommutative torus.

We refer the readers to
the textbooks \cite{connes,madore,landi,bond} for general references 
on noncommutative geometry,
\cite{GSW,polchinski,cvj} for an introduction to string theory. And to
the excellent reviews \cite{trieste,harvey,schwarz-review,nd,szabo} 
for an in-depth discussion of the physical aspects of the modern
development of noncommutative geometry in string theory, as well as for a  
more comprehensive list of references.

\head{Matrix theory compactification and noncommutative geometry}
M-theory is an eleven dimensional quantum  theory 
of gravity which is believed to underlie all superstring theories. 
Banks, Fischler, Shenker and Susskind proposed that the large
$N$ limit of the supersymmetric  matrix quantum mechanics of $N$
D0-branes should describe the $M$ theory compactified on a light-like
circle \cite{bfss}. See also \cite{taylor-M-review} for review.
Compactification of Matrix theory on torus  can be easily achieved by
considering the torus as the quotient space $\RR^d/\ZZ^d$ with the
quotient conditions
\be \label{compact}
U_i^{-1} X^j U_i = X^j + \d_i^j 2\pi R_i, \quad i = 1, \cdots, d.
\ee
Here $R_i$ are the radii of the torus. 
The unitary translation generators $U_i$  generate the torus. 
They satisfy $U_i U_j = U_j U_i$. 
T-dualizing the D0-brane system, 
the equation \eq{compact} leads to the dual description as a
$(d+1)$-dimensional supersymmetric gauge theory on the dual toroidal
D-brane \cite{taylor-compact}. A
noncommutative torus $T^d_\th$ is defined by the modified relations
\be \label{nc-T}
U_i U_j = e^{i \th_{ij}} U_j U_i,
\ee 
where $\th_{ij}$ specify the noncommutativity.  
Compactification on noncommutative torus can be easily accommodated and
leads to noncommutative gauge theory \cite{connes-rieffel}
on the dual D-brane.  
The parameters $\th_{ij}$  can be identified 
with the components $C_{-ij}$ of the 3-form potential in M-theory.  

Since M-theory compactified on a circle leads to IIA string theory, the
components $C_{-ij}$ corresponds to the Neveu-Schwarz (NS) B-field $B_{ij}$ 
in IIA string theory. The physics of the D0-brane system in the
presence of a NS $B$-field  \cite{cheung} can also be
studied from the view of IIA string theory. This lead Douglas and Hull
to obtain the same result that a noncommutative field theory lives
on the D-brane \cite{douglas-hull}. Toroidally compactified IIA string theory has a
T-duality group $SO(d,d;\ZZ)$. 
The T-duality symmetry got translated into
an equivalence relation between gauge theories on noncommutative torus
\cite{cds,schwarz-T,zumino-T}:
A gauge theory on noncommutative torus $T^d_\th$ is equivalent to 
a gauge theory on noncommutative torus $T^d_{\th'}$ if
their noncommutative parameters and metrics 
are related by the T-duality transformation. For example,
\be \label{mor}
\th' = (A\th +B)(C \th +D)^{-1}, \quad 
\left(\begin{array}{cc}A & B \\ C & D \end{array} \right) \in SO(d,d;\ZZ).
\ee
It is remarkable that the T-duality acts within the field theory level,
rather than mixing up the field theory modes with the string winding
states and other stringy excitations.
Mathematically \eq{mor} is precisely the condition 
for the noncommutative tori $T^d_\th$ and $T^d_{\th'}$ to be Morita 
equivalent \cite{connes}.

\head{Open string in $B$-field}
It was soon realized that  the D-brane is not 
necessarily to be toroidal in order to be noncommutative \cite{CH}. 
A direct canonical quantization of the open string
system shows that a constant $B$-field on a D-brane leads to
noncommutative geometry on the D-brane worldvolume. 
Consider an open
string moving in a flat space with metric $g_{ij}$ and a constant 
NS $B$-field. In the presence of a D$p$-brane, the   
components
of the $B$-field not along the brane can be gauged away and thus the
$B$-field can have effects only in the longitudinal directions along
the brane.  The worldsheet (bosonic) action for this part is 
\be \label{action} 
S= \frac{1}{ 4\pi\alpha'} \int_{\Sigma} d^2\sigma 
\bigl( g_{ij} \del_a x^i \del^a x^j - 2 \pi  \a' B_{ij} \e^{ab}\del_a
x^i \del_b x^j \bigr), 
\ee
where $i,j=0,1,\cdots, p$ is along the brane.
It is easy to see that the boundary condition 
$g_{ij} \del_\s x^j + 2 \pi i \a' B_{ij} \del_\t x^j =0$ 
at $\s=0, \pi$ is not compatible with
the standard canonical quantization $[x^i(\t,\s),x^j(\t,\s')] =0$ 
at the boundary.
Taking the boundary condition as constraints and perform the
canonical quantization, one obtains the commutation
relations
\be  \label{cr2}
[a_m^i, a_n^j]= m G^{ ij}\d_{m+n},
\quad [x_0^i, p_0^j]=i G^{ ij}, \quad
[x_0^i,x_0^j]= i \,\theta^{ij}.
\ee
Here the open string mode expansion is
\be \label{X-expan}
x^i(\tau,\sigma) =x_0^i + 2\a' (p_0^i \tau -  
2\pi \a' (g^{-1} B)^i_j p_0^j \sigma) +
\sqrt{2\a'} \sum_{n\neq 0} \frac{e^{-\ii n\tau}}{ n}
\bigl(\ii a^i_n \cos n\sigma -   2\pi \a' (g^{-1} B)^i_j a_n^j \sin
n\sigma \bigr).  \nn
\ee 
$G^{ij}$ and $\th^{ij}$ are the symmetric and 
anti-symmetric parts of the  matrix $(g+2\pi \a'B)^{-1}{}^{ij} $,
\be \label{Gth}
G^{ij} =  (\frac{1}{g+2\pi \a'B} g \frac{1}{g-2\pi \a'B})^{ij}, \quad
\th^{ij}= -(2 \pi \a')^2  
(\frac{1}{g+2\pi\a'B} B \frac{1}{g-2\pi \a' B})^{ij}.
\ee
It follows from \eq{cr2} that the  boundary coordinates 
$x^i \equiv x^i(\tau,0)$ obey the commutation relation
\be\label{boundy-cr}
[x^i,x^j] = i \th^{ij}.
\ee
The relation \eq{boundy-cr} implies that the 
D-brane worldvolume, where the
opens string endpoints live, is a noncommutative manifold. 
One may also starts with the closed string Green function and let 
its arguments to approach the boundary to obtain the open string Green
function 
\be \label{treeG}
\langle x^{i}(\t)  x^{j}(\t')  \rangle
= -\a' G^{ij} \ln (\t-\t')^2
+ \frac{i}{2} \th^{ij} \e(\t-\t'),
\ee
where $\e(\t)$ is the sign of $\t$.
From \eq{treeG},
Seiberg and Witten \cite{SW} extract the commutator \eq{boundy-cr} again.
$G_{ij} = g_{ij} - (2\pi \a')^2 (Bg^{-1} B)_{ij}$ 
is called the  open string metric since it 
controls the short distance behaviour of open strings.
In contrast, the short distance behaviour for closed strings 
is controlled by the closed string metric $g_{ij}$.
One may also treat the boundary $B$-term in \eq{action}  
as a perturbation to the open string conformal field theory and
from which one may extract \eq{treeG} 
from the modified operator product expansion of the open string vertex
operators \cite{schomerus}.

D-branes in Wess-Zumino-Witten model provides another example of
noncommutative geometry. In this case, the background is not flat since
there is a nonzero $H = dB \sim k^{-1/2}$, where $k$ is the level.
Examining the vertex operator algebra, \cite{ars} obtains that
D-branes are described by nonassociative deformation of fuzzy spheres
with nonassociativity controlled by $1/k$.

\head{String amplitudes and effective action}
The effect of the $B$-field on the open string amplitudes is
simple to determine since only the $x_0^i$ commutation
relation is affected nontrivially.
For example the {\it noncommutative gauge theory} can be 
obtained from the tree level string amplitudes readily \cite{ncsym,SW}.
For tree and one loop
\cite{1loop},  
we can use
the vertex operator formalism. Generally the vertex operator can
be inserted at either the $\s=0$ or $\s=\pi$ border, where the string
has the zero mode parts $x_0^i$ and  
$y_0^i \equiv x_0^i - (2\pi \a')^2 (g^{-1} B)^i_j p_0^j$ 
respectively.  The commutation relations are
\be \label{xy-cr}
[x_0^i,x_0^j] = i \th^{ij},\quad
[x_0^i,y_0^j] =0,\quad [y_0^i,y_0^j] = - i \th^{ij}.
\ee
The difference in the commutation relation for $x_0$ and $y_0$
implies that the two borders of the open string has opposite
commutativity. This fact
is not so important for tree level calculations 
since one can always choose to put all the interactions at, for
example,  the  $\s=0$ border. Collect all these zero mode parts of the
vertex operators, one obtain a phase factor 
\be \label{phase}
e^{i p^1 x_0} e^{i p^2 x_0} \cdot  e^{i p^N x_0} = 
e^{i \sum p^a x_0} e^{ -\frac{i}{2} \sum_{I<J}^N p^I \th p^J},
\ee
where the external momenta $p^a$ are ordered cyclically on the circle
and  momentum conservation has been used. 
The computation of the oscillator part of  the amplitude is the same 
as in the $B=0$ case, except that the metric $G$ is employed in 
the contractions.  As a result, the effect of $B$-field on the tree
level string amplitude is simply to   
multiply the amplitude at $B=0$ with the  phase factor and
to replace the metric by the metric $G$. 
A generic term in the tree level effective action simply becomes
\be \label{replace}
\int d^{p+1} x \sqrt{-\det g}\; \tr \del^{n_1} \Phi_1  \cdots 
\del^{n_k} \Phi_k
\to \int d^{p+1} x \sqrt{-\det G}\; \tr \del^{n_1} \Phi_1 * \cdots 
 * \del^{n_k} \Phi_k.
\ee
Here the star-product, also called the Moyal product is defined by
\be
(f * g) (x) = e^{i \frac{\th^{\mu\nu}}{2} 
\frac{\del}{\del x_1^\mu} \frac{\del}{\del x_2^\nu} 
} f(x_1) g(x_2) |_{x_1=x_2}.
\ee
The star-product is associative,  noncommutative and satisfies 
$\overline{f * g} = \bar{g} * \bar{f}$
under complex conjugation. Also for functions vanish rapidly enough at infinity, 
it is 
\be \label{integ}
\int f*g = \int g*f = \int f g.
\ee
An interesting consequence of the nonlocality 
as expressed by the noncommutative geometry \eq{boundy-cr} 
is the existence of dipole excitation \cite{dipole} 
whose extent is proportional to its momentum, $\Delta x = k \th$ . 
This relation is at the heart of the {\it IR/UV mixing phenomena} 
(see below) of noncommutative field theory. 
Moyal product in string theory has also appeared earlier in the works 
\cite{earlier}.

At  one (and higher) loop level, the 
different noncommutativity for the opposite boundaries of open
string becomes essential and give rises to new effects \cite{reggeon}. 
In this case non-planar
diagrams require to put vertex operators at the two different
borders $\s=0,\pi$. A more complicated phase factor 
which involves internal as well as external momentum is resulted.
This leads to the 
IR/UV mixing in the noncommutative quantum
field theory. The  
different noncommutativity for the opposite boundaries of open
string \eq{xy-cr} is the basic reason  for the  IR/UV
mixing in the noncommutative quantum field theory.
The commutation relations \eq{cr2} are valid at all loops and
therefore one can use them to construct the higher loop
string amplitudes from the first principle \cite{reggeon}. The effect of the
$B$-field on the string interaction can be easily implemented 
into the Reggeon vertex and the complete higher loop amplitudes 
in B-field have been constructed.

\head{Low energy limit: the Seiberg-Witten limit and the NCOS limit}
The full open string system is still quite complicated.  
One may try to decouple the infinite number of massive 
string modes to obtain a low energy
field theoretic description by taking the limit $\a' \to 0$. 
Since open string are sensitive to $G$ and $\th$,
one should take the limit such that $G$ and $\th$ are fixed.
For the magnetic case $B_{0i} =0$, 
Seiberg and Witten \cite{SW} showed that  this can be achieved with the 
following double scaling limit
\be
\a' \sim \e^{1/2}, \quad g_{ij} \sim \e \to 0
\ee
with $B_{ij}$ and everything else  kept fixed. Assuming $B$ is of rank
$r$, then \eq{Gth} becomes
\be \label{Gth-limit}
G_{ij} = -(2\pi \a')^2 (B g^{-1} B)_{ij},\quad
\th^{ij} = (B^{-1})^{ij}, \quad \mbox{for} \;\; i,j = 1,\cdots, r.
\ee
Otherwise $G_{ij} = g_{ij}$, $\th^{ij} =0$. One may also argue that 
closed string decouples in this limit.  
As a result, we obtain  at the low energy limit 
a greatly simplified noncommutative Yang-Mills action $F*F$. 
See below for more discussion of this field theory. 

For the case of a constant  electric field background, say $B_{01}
\neq 0$, there is a critical electric field beyond which open string 
become unstable and the theory does not make sense. 
Due to the presence of this upper
bound of the electric field, one can show that there is no decoupling
limit where one can reduce the string theory to a field theory on a
noncommutative spacetime.  However one can consider a different
scaling limit where one takes the closed string metric is
scale to infinity appropriately as the electric field approaches the
critical value.  In this limit, all  closed string modes 
decoupled. One obtains a novel noncritical string theory  
living on a noncommutative spacetime 
known as Noncommutative Open String (NCOS). For more details, see
\cite{ncos1,ncos2}

\section{Noncommutative Quantum Field Theory}
 
Field theory on noncommutative spacetime are defined using the star-product
instead of ordinary product of the fields. To illustrate the general
ideas, let us consider a single real scalar field theory with the action
\be
S  = \int d^D x \left[ \frac{1}{2} (\del \phi)^2 - \frac{m^2}{2} \phi^2 -
V(\phi) \right], \quad V(\phi) = \frac{g}{ 4!} \phi^{*4}.
\ee
Due to the property \eq{integ},  free noncommutative field theory is the
same as an ordinary field theory. Treating the interaction term as 
perturbation, one can perform the usual quantization and obtain the 
Feynman rules: the propagator is unchanged and the interaction vertex in
the momentum space is given by $g$ times the phase factor
\be \label{vtx}
 \exp \left(- \frac{i}{ 2} \sum_{1\leq a<b\leq 4} p^{a} \times p^{b}
\right).
\ee
Here $p\times q \equiv p_\m \th^{\m\n} q_\n$. The theory is nonlocal due
to the infinite order of derivatives that appear in the interaction. 

\head{Planar and non-planar diagrams} The factor \eq{vtx} is 
cyclically symmetric but not permutation symmetric. This is analogous to
the situation of a matrix field theory. Using the same double line
notation as introduced by 't Hooft, one can similarly classify the Feynman
diagrams of noncommutative field theory according to its genus. In
particular the total phase factor of a planar diagram behaves quite
differently from that of a non-planar diagram. 
It is easy to 
show that a planar diagram will have the phase factor 
\be \label{planar-V}
V_{\rm p}(p^1, \cdots, p^n) = \exp \left( - \frac{i}{ 2} 
\sum_{1\leq a<b\leq n} p^{a} \times p^{b}
\right) ,
\ee
where $p^1, \cdots, p^n$ are the (cyclically ordered) 
external momenta of the graph. Note that the phase factor \eq{planar-V} is
independent of the internal momenta. This is not the
case for a non-planar diagram. One can easily show that a non-planar diagram 
carries an additional phase factor \cite{filk}
\be\label{nonplanar-V}
V_{\rm np} = V_{\rm p} \exp \left(  - \frac{i}{ 2} 
\sum_{1\leq  a<b\leq n} C_{ab} p^{a} \times p^{b}
\right),
\ee
where the $C_{ab}$ is the signed intersection matrix of the graph, 
whose $ab$ matrix element counts  the number of times the $a$-th
(internal or external) line crosses the $b$-th line. The matrix $C_{ab}$
is not uniquely determined by the diagram as different ways of drawing
the graph could lead to different intersections. However the phase factor 
\eq{nonplanar-V} is unique due to momentum conservation.
  
This different behaviour of the planar and non-planar phase factor has 
important consequences.

{\tt 1.} Since the phase factor \eq{planar-V} is independent of the 
internal momenta,  the divergences and renormalizability of the 
planar diagrams will be (simply) the same as in the commutative theory
and can be handled with standard renormalization techniques.
This is sharply different for the non-planar diagrams. In fact due to the 
extra oscillatory internal 
momenta dependent phase factor, one can expect the non-planar diagrams
to have an improved ultraviolet (UV) behaviour. It turns out 
that planar and non-planar diagrams also differ sharply in their 
infrared (IR) behaviour due to 
the IR/UV mixing effect. 

{\tt 2.} Moreover at  high energies, one can expect 
noncommutative field theory will 
generically  become planar since the non-planar diagrams will be 
suppressed  due to the oscillatory phase factor.

{\tt 3.} In the limit $\th\to\infty$, 
the non-planar sector will be totally suppressed since the
rapidly oscillating phase factor will cause the non-planar 
diagram to vanish upon integrating out the momenta.
Thus generically the large $\th$ limit 
is analogous to the large $N$ limit where only the 
planar diagrams contribute.  However these 
expectations do not apply for noncommutative gauge theory since one need to
include {\it open Wilson line} (see below) in the construction of gauge invariant 
observables, and the open Wilson line grows in extent with energy and $\th$. 
 
\head{IR/UV mixing}
Due to the nonlocal nature of noncommutative field theory, there 
is generally  a 
mixing of  the UV and IR scales \cite{iruv1}. 
The reason is roughly the follow. 
Non-planar diagrams generally has phase factors like 
$\exp (i k\th p)$ , with $k$ a loop momentum, $p$ an external momentum. 
Consider 
a non-planar diagram which is UV divergent when $\th=0$,
one can expect that for very 
high loop momenta, the phase factor will oscillate rapidly and 
renders the integral finite.  However this is only valid for a
non-vanishing external momentum $\th p$, the infinity will come back as 
$\th p \to 0$. However this time it appears as an IR singularity. Thus
an IR divergence arises
whose origin is from the UV region of the momentum integration and this
is known as the IR/UV mixing phenomena.

To be more specific, consider the $\phi^4$ scalar theory in $D=4$ 
dimensions. 
The one loop self energy has a non-planar contribution given by  
\be \label{two-point}
\Gamma_{\rm np} = \frac{g}{6 (2\pi)^4} \int \frac{d^4 k}{ k^2 +m^2} 
e^{i k\th p} \sim \frac{g}{3 (4\pi^2)^{2}} (\L_{\rm eff}^{2} + \cdots),
\ee
where $\L_{\rm eff}^2 = (1/ \L^2 + (\th p)^2)^{-1}$.
One can see clearly the IR/UV mixing:
$\Gamma_{\rm np}$ is UV finite as long as $\th p \neq 0$;
when $\th p =0$, the quadratic UV divergence is recovered, 
$\Gamma_{\rm np} \sim \L^{2}$. For supersymmetric theory, 
we have at most logarithmic IR 
singularities from IR/UV mixing.

IR/UV mixing has a number of interesting consequences. 

{\tt 1.} Due to the IR/UV mixing, noncommutative theory does not appear 
to have a consistent Wilsonian  description since it will require that 
correlation functions computed at finite $\L$
differ from their limiting values by terms of order $1/\L$ for all
values of momenta. However this is not true for theory with IR/UV
mixing. For example, the two point function \eq{two-point} at finite
value of $\L$ differs from its value at $\L =\infty$ by the amount 
$ \Gamma_{\rm np}^\L - \Gamma_{\rm np}^{\L = \infty}
\propto 1/(\th p)^2$,
for the range of momenta $(\th p)^2 \ll 1/\L^2$. It has been 
argued that  the IR singularity may be associated with
missing light degrees of freedom in the theory. With new degrees of 
freedom appropriately added, one may
recover a conventional Wilsonian description. Moreover it 
has been suggested
that to identify these degrees of freedom with the closed string modes
\cite{iruv1}.
However the precise 
nature and origin of these degrees of freedom is not  known.

{\tt 2.} The renormalization of the planar diagrams is straight
forward, however it is more subtle for the non-planar diagrams since
the IR/UV-mixed IR singularities may mix with other divergences
at higher loops and render the proof of renormalizability much 
more difficult.  
IR/UV mixing renders certain large $N$ noncommutative field theory
nonrenormalizable \cite{largeN}. 
However for theories with a fixed set of degree of freedoms to start with,
it is believed that one can have sufficiently good control of the IR 
divergences and prove renormalizability. 
An example of renormalizable noncommutative
quantum field theory is the noncommutative  Wess-Zumino model \cite{wzw} 
where IR/UV mixing is absent.
However a general proof is still lacking despite the progress made in
\cite{renorm}. 

{\tt 3.} One can show that IR/UV mixing in timelike noncommutative  
theory ($\th^{0i} \neq 0$) leads to breakdown of perturbative 
unitarity \cite{u1,u2,u3}. 
For theory without IR/UV mixing, unitarity will be respected even if 
the theory has a timelike noncommutativity \cite{u3}. Theory with
lightlike noncommutativity is unitary \cite{u4}.

\section{Noncommutative Gauge Theory}
  
Gauge theory on noncommutative space is defined by the action
\be \label{ncym}
S =-\frac1{4g^2}\,\int d x~\tr\Bigl(F_{ij}(x)*  F^{ij}(x)\Bigr) ,
\ee
where the gauge fields $A_i$ are $N\times N$ Hermitian matrices,
$F_{ij}$  is the noncommutative field strength 
$ F_{ij}=\partial_i A_j-\partial_j A_i-i [A_i, A_j]_* $
and $\tr$ is the ordinary trace over $N\times N$ matrices.
The theory is invariant under the star-gauge transformation
\be \label{nct1}
 A_i \to g*A_i*g^\dagger - i g * \del_i g^\dagger,
\ee
where the $N\times N$ matrix function $g(x)$ is unitary with respect
to the star-product $ g* g^\dagger = g^\dagger * g = I$. The solution is
$g =e_*^{i \l}$ where $\l$ is Hermitian.  
In infinitesimal form, 
$ \d_\l A_i = \del_i \l + i [\l, A_i]_*$ .
The noncommutative gauge theory has $N^2$ Hermitian gauge fields. 
Because of the star-product, the $U(1)$ sector of the theory is not free 
and does not decouple from the $SU(N)$ factor as in the
commutative case.  
Note that this way of defining noncommutative gauge theory does not work
for the other Lie group since the star-commutator generally involves
commutator as well as anti-commutator of the Lie algebra, and hence the 
expressions above  generally involve the enveloping algebra 
of the underlying Lie group. With the help of the 
{\it Seiberg-Witten map} (see below), 
one can  construct an enveloping algebra valued gauge theory
which has the same number of independent gauge fields and gauge
parameters as the ordinary Lie algebra valued gauge theory \cite{wess}. 
However the quantum properties of these theories are much less
understood.
One may also introduce certain automorphism in the 
noncommutative $U(N)$ theory 
to  restrict the dependence of the
noncommutative space coordinates of the field configurations and obtain a 
notion of noncommutative theory with orthogonal and symplectic 
star-gauge group \cite{sosp}.  However the theory does not reduce to the
standard gauge theory in the commutative limit $\th \to 0$. 

\head{Open Wilson line and gauge invariant observables}
One remarkable feature of noncommutative gauge 
theory is the mixing of  noncommutative
gauge transformation and spacetime translation, as can be seen from
the following identity
\be \label{transl-prop}
e^{i kx} * f(x) * e^{-ikx} = f(x+ k\theta),
\ee
for any function $f$. This is analogous to the situation in general
relativity where translations are also equivalent to 
gauge transformations (general
coordinate transformations). Thus as in general relativity, 
there are no local gauge invariant
observables in noncommutative gauge theory.
The unification of spacetime and gauge fields in 
noncommutative gauge theory can also be seen from the fact that
derivatives can be realized as commutator 
$\del_i f \to -i [ \th^{-1}_{ij} x^j, f]$
and get absorbed into the vector potential in
the covariant derivative
\be \label{cov-D}
D_i  = \del_i + i A_i \to  -i  \th^{-1}_{ij} x^j + i A_i.
\ee
Equation \eq{cov-D} clearly demonstrates the unification of spacetime and
gauge fields. Note that the field strength takes the form 
$F_{ij} = i [D_i,D_j] + \th^{-1}_{ij}$. 

The Wilson line operator for a path $C$
running from $x_1$ to $x_2$ is defined by
\be
W(C) = 
P_* \exp(i \int_{C} A). 
\ee
$P_*$ denotes the 
path ordering with respect to the star-product, with $A(x_2)$ at
the right. It transforms as 
\be \label{W-transf}
W(C) \to g(x_1)*W(C)*g(x_2)^\dagger.
\ee
In commutative gauge theory, 
the Wilson line operator for closed loop (or its Fourier transform) 
is gauge invariant.
In noncommutative gauge theory, the closed Wilson loops are no longer gauge
invariant.  Noncommutative 
generalization of the  gauge invariant Wilson loop operator can be
constructed most readily by deforming the  Fourier transform of the 
Wilson loop operator. 
It turns out that the closed loop has to open in a specific way to
form an open Wilson line in order to be gauge invariant \cite{wilson-line}.
To see this, let us consider a path $C$ 
connecting points 
$x$ and $x +l$. Using \eq{transl-prop}, it is easy to see that 
the  operator
\be \label{open-w}
\tilde{W}(k) \equiv \int dx\; \tr W(C) *  e^{i k x}, \quad  
\mbox{with} \quad l^{j}=k_i \theta^{ij},
\ee
is gauge invariant. 
Just like Wilson loops in ordinary gauge theory, these operators also
constitute an over complete set of gauge invariant operators 
parametrized by the set of curve $C$. When $\th=0$, $C$
become a closed loop and we re-obtain the (Fourier transform) 
usual closed Wilson loop in commutative gauge theory. 
Noncommutative version of the loop equation for closed Wilson loop  
has been constructed and involves open Wilson line \cite{loopeqn}. 
The open Wilson line is instrumental in the construction of gauge invariant observables 
\cite{rey-w1,gross-obs,rey-w2,DW}. An important application  is in the construction 
of various couplings of
the noncommutative D-brane to the bulk supergravity fields 
\cite{sugra13}. The
equivalence of the commutative and noncommutative couplings to the
RR-fields has leaded to
the exact expression for the Seiberg-Witten map \cite{liu-exact}. 
It is remarkable that the one-loop nonplanar effective action for
noncommutative scalar theory \cite{eff-s1}, gauge theory 
\cite{eff-g1a,eff-g1b,eff-g1c}
as well as the two-loop effective action for scalar \cite{eff-s2} 
can be written compactly in terms of open Wilson line.
Based on this result, the physical origin of the IR/UV mixing has been
elucidated. One may identify the  open Wilson line
with the dipole excitation generically presents 
in noncommutative field theory \cite{eff-s1,eff-s2} and hence
explain the presence of the IR/UV mixing.
IR/UV mixing may also be identified with 
the instability \cite{eff-g1a,eff-g1b} associated
with the closed string exchange of the noncommutative D-branes.

\head{The Seiberg-Witten map} 
The open string is coupled to the one-form $A_i$ living on the
D-brane through the coupling $\int_{\del \S} A$.  For slowly varying fields, 
the effective action for this gauge potential can be determined from
the S-matrix and is given by the
Dirac-Born-Infeld (DBI) action. In the presence of $B$-field, the discussion
above \eq{replace} leads to the noncommutative DBI Lagrangian
\be \label{DBI1}
L_{NCDBI}(\hat{F}) = G_s^{-1} \mu_p \sqrt{- \det (G + 2\pi \a' \hat{F})} ,
\ee
where $\mu_p = (2\pi)^{-p} (\a')^{-(p+1)/2}$ is the D-brane tension
and $\hat{F}$ is the noncommutative field strength.
However one may also exploit the tensor gauge invariance on the 
D-brane (that is, the string sigma model is invariant under 
$A \to A- \L, B \to B + d \L$) and consider
the combination $F+B$ as a whole. In this case it is like having the
open string coupled to  the boundary gauge field strength $F+B$ and
there is no $B$ field. And we have the usual DBI Lagrangian \cite{DBI12}
\be \label{DBI2}
L_{DBI}(F) = g_s^{-1} \mu_p  \sqrt{- \det (G + 2\pi \a' (F+B))} .
\ee
In \eq{DBI1} and \eq{DBI2}, $G_s$ and $g_s$ are the effective open 
string couplings in the noncommutative and commutative descriptions.
Although looks quite different, Seiberg and
Witten showed the commutative and noncommutative DBI actions 
are indeed equivalent if the open string couplings are related by
$g_s =G_s \sqrt{\det (g+ 2\pi \a' B)/\det G} $
and there is a field redefinition that relate the 
commutative and noncommutative gauge fields. The map 
$\hat{A} = \hat{A} (A)$ is called the Seiberg-Witten map \cite{SW}.
Moreover, the noncommutative gauge symmetry is equivalent to the 
ordinary gauge symmetry in the  sense that they have the same set of 
orbits under gauge transformation:
\begin{equation}
\label{swmap-def}
\hat A(A) + \hat\delta_{\hat\l} \hat A(A) =
\hat A(A + \delta_\l A) .
\end{equation}
Here $\hat A_i$ and  $\hat\l$ are the  noncommutative gauge field and  
noncommutative gauge transformation parameter,   
$A_i$ and  $\l$ are the  ordinary gauge field and  ordinary 
transformation parameter.  
The map between $\hat A_i$ and $ A_i$ is called the Seiberg-Witten map. 
\eq{swmap-def} can be solved only if the transformation parameter
$\hat\l = \hat \l (\l, A)$ is field dependent. 
The Seiberg-Witten map is characterized by the Seiberg-Witten 
differential equation, 
\be
\delta \hat A_i(\theta ) = -\frac{1}{4}\delta\theta^{kl} \big[\hat A_k*
(\partial_l \hat A_i + \hat F_{li}) + (\partial_l \hat A_i +\hat
F_{li})* \hat A_k \big] .
\ee
An  exact solution for the Seiberg-Witten map can be written down with
the help of the  open Wilson line.
For the case of $U(1)$  with constant $F$, we have the exact 
solution
$ \hat{F} = (1+ F \th)^{-1} F$.

That there is a field redefinition that allows one to write the
effective action in terms of different fields with different gauge
symmetries may seems puzzling in the first instant. However it has a
clear physical origin in terms of the string worldsheet. In fact there
is different possible schemes to regularize the short distance
divergence on the worldsheet. One can show that the Pauli-Villar
regularization gives the commutative description, while the point
splitting regularization gives the noncommutative description. Since
theories defined by different regularization schemes are related by a
coupling constant redefinition, this implies the commutative and
noncommutative description are related by a field redefinition because
the couplings on the worldsheet are just the spacetime fields. 

Despite this formal equivalence, the physics of the noncommutative
theories is generally quite different from the commutative one.
First it is clear that generally the Seiberg-Witten map may  
take nonsingular configurations to singular configurations.
Second, the observables one is interested in are also generally different. 
Moreover the two descriptions are generally good for different regimes:
the conventional gauge theory description is simpler for small $B$ and 
noncommutative  description is simpler for large $B$.

\head{Perturbative gauge theory dynamics}
The noncommutative gauge symmetry \eq{nct1} 
can be fixed as usual by employing the
Fadeev-Popov procedure, resulting in Feynman rules that are similar to
the conventional gauge theory. The important difference is that 
now we have to
amend to the structure constant the phase factors
\eq{planar-V} and \eq{nonplanar-V}.
It turns out that the non-planar
$U(N)$ diagrams contribute (only) to the $U(1)$ part of the theory. As a
result, unlike the commutative case, the $U(1)$ part of the theory 
is no longer decoupled and free \cite{armoni}.  
Noncommutative gauge theory is one-loop renormalizable \cite{renorm-gauge,renorm-gaugeN}. 
The beta-function is determined
solely by the planar diagrams and, at one loop, is given by 
\cite{renorm-gauge,renorm-gaugeN,iruv2}
\be
\beta(g) = -\frac{22}{ 3} \frac{N g^3}{16 \pi^2}, \quad
\mbox{for $N \geq 1$}.
\ee
Note that the beta function is independent of $\th$,  
the noncommutative $U(1)$ is asymptotically free and 
does not reduce to the commutative theory when $\th \to 0$.
Noncommutative theory beyond  the tree level
is generally not smooth in the limit $\th \to 0$. 
Discontinuity of this kind was also noted for the Chern-Simon system 
\cite{CS}.

Gauge anomalies can be similarly discussed \cite{anom1,anom2,anom3} and satisfy the
noncommutative generalizations of the Wess-Zumino consistency
conditions \cite{anom3}. In $d=2n$ dimensions, the anomaly involves the combination
$\tr (T^{a_1} T^{a_2} \cdots T^{a_{n+1}})$ rather than the usual symmetrized trace
since the phase factor is 
not permutation symmetric. As a result, the usual 
cancellation of anomaly does not work and is the main obstacle to the
construction of noncommutative chiral gauge theory.

There is a number of interesting features to mention for the
IR/UV mixing \cite{iruv2,iruv3}  in noncommutative gauge theory:
{\tt 1.} IR/UV mixing generically yields pole like  IR
singularities. Despite the appearance of IR poles,  
gauge invariance of the theory is not endangered \cite{iruv3}. 
{\tt 2.} One can show that only the $U(1)$ sector is affected 
by IR/UV mixing.
{\tt 3.} As a result of IR/UV mixing,  noncommutative $U(1)$ photons 
polarized in the noncommutative 
plane will have different dispersion relations from those which are not \cite{iruv3}.
Strange as it is, this is consistent with gauge invariance.

\section{Noncommutative Solitons, Instantons and D-branes}

Solitons and instantons play important roles in the nonperturbative
aspects of field theory. The nonlocality of star-product gives noncommutative
field theory a stringy nature.
It is remarkable that this applies to the nonperturbative sector as well.
Solitons and instantons in the noncommutative gauge theory 
reproduce amazingly the properties of D-brane in the string. The reviews
\cite{trieste, harvey} are recommended.

\head{GMS solitons}
Derrick theorem says that commutative scalar field theories in 
two or higher dimensions 
do not admit any finite energy classical solution. This follows from a
simple scaling argument, which will fail when the theory becomes 
noncommutative 
since noncommutativity introduce a fixed length scale $\sqrt{\th}$.
Noncommutative solitons in pure scalar theory can be easily constructed
at the limit $\th = \infty$. For example, consider a (2+1)-dimensional 
single scalar theory with a potential $V$ and noncommutativity 
$\th^{12} =\th$. In the limit $\th = \infty$, 
the potential term dominates and 
the noncommutative solitons are determined by the equation
\be \label{Vs}
\del V / \del \phi =0.
\ee
Equation \eq{Vs} can be easily 
solved in terms of projector. Assuming $V$ has no linear term, 
the general soliton (up to unitary equivalence) is
\be\label{nc-soliton}
\phi = \sum_i \l_i P_i,
\ee
where $\l_i$ are the roots of $V'(\l) =0$ and 
$P_i$ is a set of orthogonal projectors. For
real scalar field theory, we have to restrict the sum to real roots only.
These solutions are known as the Gopakumar-Minwalla-Strominger (GMS) 
solitons \cite{gms}.
A simple example of projector is  given by $P =  \ket{0}\bra{0}$, 
which corresponds to a Gaussian profile in the $x^1, x^2$ plane with width 
$\sqrt{\th}$. The soliton continues to exist until $\th$ decreases below a
certain critical $\th_c$.

New solution can be generated from known one using the so called  
solution generating technique.
If $\phi$ is a solution of \eq{Vs}, then 
\be
\phi' = T^\dagger \phi T
\ee
is also a solution provided that $T T^\dagger =1$.
In an infinite dimensional Hilbert space, $T$ is not necessary to be
unitary, i.e. $T^\dagger T \neq 1$.
In this case, $T$ is said to be a partial isometry. The
new solution $\phi'$ is different from $\phi$ since they are
not related by  global transformation of basis.
 
\head{Tachyon condensation and D-branes}
A beautiful application of the noncommutative soliton is in the
construction of D-brane as solitons of the tachyon field in
noncommutative open string theory 
\cite{soliton-d}
For the bosonic string theory, one may consider it a space filling
D25-brane. Integrating out the massive string modes lead to an effective
action for the tachyon and the massless gauge field $A_\m$. 
We remark that different from the pure scalar case, 
noncommutative soliton can be constructed exactly for finite $\th$ in a
system with gauge and scalar fields \cite{ncsol-exact}.  
Although the detailed form of the effective action is unknown, 
one has enough confidence to say what the true vacuum configuration is 
according to the Sen conjecture. See for example, \cite{TZ-rev} 
for excellent review
on this subject. One can then apply the 
solution generating technique to generate new soliton solution.
With a $B$-field of rank $2k$, one can construct this way solutions which are
localized in $\RR^{2k}$ and represents a D$(25-2k)$-brane. This is 
supported by the  matching of the tension and the spectrum of
fluctuation around the soliton configuration. 
Similar ideas can also be applied to construct D-branes in type II
string theory. Again the starting point is an unstable brane configuration
with tachyon field(s). There are two types of unstable D-branes:
non-BPS D$p$-branes ($p$ odd in IIA theory and $p$
even in IIB theory) and 
BPS branes anti-branes D$p$-$\overline{{\rm D}p}$ system.
A similar analysis 
allows one to identify the noncommutative soliton with the 
lower dimensional BPS D-branes which arises from tachyon condensation.

One main motivation for studying tachyon condensation in open string
theory is the hope that open string theory may provide a fundamental
nonperturbative formulation of string theory. It may not be too
surprising that D-branes can be obtained in terms of open string fields.
However to describe closed string and NS-branes in terms of open string
degree of freedom remain an obstacle.

\head{Noncommutative instanton and monopoles}
Instantons on noncommutative $\bf{R}_\th^4$ can be constructed readily
using the ADHM formalism by  modifying the 
ADHM constraints with a constant additive term \cite{ns-ins}. 
The result is that
the self-dual (resp. anti-self-dual) instanton moduli space depends only on the  
anti-self-dual (resp. self-dual) part.
The construction go through even in the $U(1)$ case. Consider a self-dual
$\th$, the ADHM constraints for the 
self-dual instanton is the
same as in the commutative case, and there is no  nonsingular
solution. On the other hand, the ADHM constraints for the
anti-self-dual instanton get modified and admit nontrivial solution. This
noncommutative  instanton solution is nonsingular with size
$\sqrt{\th}$.  Although ADHM method does not  
give a self-dual instanton, a direct construction can be applied to
obtain  non ADHM self-dual instantons. Recall that the
gauge field strength can be written as 
$F_{ij} = i [D_i,D_j] + \th^{-1}_{ij}$, 
where $D_i$ is given by the function on the right hand side of \eq{cov-D}. 
Thus a simple self-dual solution can be constructed with
\be
D_i =  i \th^{-1}_{ij} T^\dagger x^j T ,
\ee
where $T$ is a partial isometry which satisfies $T T^\dagger =1$, but 
$T^\dagger T =1 -P$ is not necessarily the identity. It is clear
that $P$ is a projector. The field strength 
\be
F_{ij} = \th^{-1}_{ij} P
\ee
is self-dual and has an instanton number $n$ where $n$ is the rank of
the projector. 
The noncommutative instanton represent a D$(p-4)$-brane within a D$p$-brane.
The ADHM constraints are just the D-flatness condition for 
the D-brane worldvolume  gauge theory.
The additive constant to the ADHM constraints also has a simple
interpretation as a Fayet-Illiopolous parameter which appears 
in the presence of $B$-field.
 
On noncommutative $\RR^3$  (say $\th^{12}= \th$), 
BPS monopoles satisfy the Bogomolny equation \cite{mono}
\be
\nabla_i \Phi = \pm B_i, \quad i =1,2,3
\ee
and can be obtained by solving the Nahm equation
\be
\del_z T_i = \e_{ijk} T_j T_k + \d_{i3} \th.
\ee
$T_i$ are  
$k\times k$ Hermitian matrices depending on an auxiliary variable 
$z$ and  $k$ gives the charge of the monopole. 
Noncommutativity modifies the Nahm equation 
with a constant term, which can be absorbed by a constant 
shift of the generators. Therefore unlike the case of instanton, the
monopole moduli space is not modified by noncommutative deformation. 
The Nahm construction has clear physical meaning in string theory. 
The monopole (electric charge) 
can be interpreted as a D-string (fundamental string) ending on a D3-brane.
One can also suspend $k$ D-string between a collection of $N$ parallel
D3-branes, this would correspond to a charge $k$ monopole in a 
Higgsed $U(N)$ gauge theory. The matrices $X^i$ correspond to the 
matrix transverse coordinates of the D-strings which lie within the 
D3-branes. 

\section{ Further Topics} 

Finally we include a brief discussions of some further topics of
interests. 

\noindent{\sf 1.} The  noncommutative geometry we discuss here is
of canonical type. Other deformations exists, for example,
kappa-deformation \cite{kappa} and fuzzy sphere which are of the Lie-algebra type,
quantum group deformation \cite{QG} which is a quadratic type deformation:
$ x^i x^j = q^{-1} \hat{R}^{ij}_{kl} x^k x^l$, whose
consistency is guaranteed by the Yang-Baxter equation. It is
interesting to see whether these noncommutative geometries arise from
string theory. Another natural generalization is to consider
noncommutative geometry of superspace. A simple example is to consider 
the fermionic coordinates to be deformed with the nonvanishing relation
\be \label{thth}
\{\th^\a, \th^\b\} = C^{\a\b}, 
\ee
where $C^{\a \b}$ are constants. It has been shown that \eq{thth} 
arises in certain Calabi-Yau compactification of type IIB string theory 
in the presence of RR-background \cite{gp13}. 
The deformation \eq{thth} reduces the number of
supersymmetries to half. Therefore it is called $\cN =1/2$ supersymmetry 
\cite{seiberg-ncsusy}.
The noncommutativity \eq{thth} can be implemented  on the
superspace $(y^i, \th^\a, \bar{\th}^{\dot{\a}})$ as a star-product
for the $\th^\a$'s. Unlike the bosonic
deformation which involves an infinite number of higher derivatives, 
the star-product for \eq{thth} stops at order $C^2$ due 
to the Grassmanian nature of the fermionic  coordinates. 
Field theory with $\cN=1/2$ supersymmetry is local and differs 
from the ordinary $\cN=1$ theory by only a 
few number of  supersymmetry breaking terms.
The $N=1/2$ WZ model is renormalizable \cite{ncsusy-wz} if 
extra $F$ and $F^3$ terms are added to the original Lagrangian, where
$F$ is the auxiliary field. 
The $N=1/2$ gauge theory is also renormalizable \cite{ncsusy-g}.

\noindent{\sf 2.} Integrability of a theory provides valuable
information beyond the perturbative level.  An integrable field theory
is characterized by an infinite number of conserved charges in involution. 
It is natural to ask whether integrability is
preserved by noncommutative deformation.  
Integrable noncommutative integrable field theories
have  been constructed, see \cite{int-field}. 
In the commutative case, Ward \cite{ward} has conjectured that all (1+1) and
(2+1)-dimensional integrable system can be obtained from the  
4-dimensional self-dual Yang-Mills equation by reduction.
Validity of the noncommutative version of the Ward conjecture
has been confirmed so far \cite{int-field}, \cite{lax}.  
It is interesting to see whether it is true in general.

\noindent{\sf 3.} Locality and Lorentz symmetry form the corner stones
of quantum field theory and standard model physics of particles.
Noncommutative field theory
provides a theoretical framework where one can discuss about effects of
nonlocality and Lorentz symmetry violation. Possible phenomenological
signals \cite{phen} have been investigated (mostly at the tree
level) and bound has been placed on the extent of noncommutativity.
Proper understanding and a better control of the IR/UV mixing remains
the crux of the problem. Noncommutative geometry may also be relevant for
cosmology and inflation \cite{cosmo}.

\noindent{\sf 4.} Like the standard AdS/CFT correspondence \cite{adscft}, the
noncommutative gauge theory should also has a gravity dual description 
\cite{ncadscft,ncadscft2}. The
supergravity background can be determined by considering the
decoupling limit of D-branes with a NS B-field background.
However since the noncommutative gauge theory does not permit
any conventional local gauge invariant observable, the
usual AdS/CFT correspondence that relates field theory correlators with
bulk interaction does not seem to apply.
It has been argued that \cite{rey-w2} generic properties like the
dipole relation between length and momentum for open Wilson line can be seen
from the gravity side. A more precise understanding of the duality
map is called for.



\noindent{\large \bf Keywords: }
Anomaly cancellation,
BFSS Matrix model,
D-branes,
M-theory,
IR/UV mixing,
Morita equivalence,
Noncommutative gauge theory,
Noncommutative geometry,
Noncommutative instantons and monopoles,
Noncommutative tori,
Sen conjecture,
Seiberg-Witten map,
Seiberg-Witten decoupling limit,
Tachyon condensation.

\noindent {\large \bf See also: } 
Superstring theories. 
Brane construction of gauge theories. 
Dualities in string theories. 
String field theory. 
String as large N-limit of gauge theory. 
Noncommutative tori, Yang-Mills and string theory. 
Deformation quantization. 
Soliton and other extended field configurations.


\begin{thebibliography}{99}

\bibitem{witten1}
E.~Witten,
``Noncommutative Geometry And String Field Theory,''
Nucl.\ Phys.\ B {\bf 268} (1986) 253.

\bm{cds}
A.~Connes, M.~R.~Douglas and A.~Schwarz,
``Noncommutative geometry and matrix theory: Compactification on tori,''
JHEP {\bf 9802}, 003 (1998)
[arXiv:hep-th/9711162].

\bm{connes}
A. Connes, ``Noncommutative geometry,'' 
Academic Press, 1994.

\bm{madore}
J. Madore, ``An Introduction to
Noncommutative Geometry and its Physical Applications,'' 
Cambridge University Press, 1999.

\bm{landi}
G. Landi, ``An Introduction to Noncommutative Spaces and their
Geometries,'' 
Springer-Verlag, 1997.

\bm{bond}
J.M. Gracia-Bond\'{\i}a, J.C. V\'arilly and H. Figueroa, 
``Elements of Noncommutative Geometry,'' 
Birkh\"auser, 2000.

\bm{GSW}
M.B. Green , J.H. Schwarz and E. Witten, ``Superstring theory,''
Cambridge Univ. Press, 1987. 

\bm{polchinski}
J. Polchinski, ``String Theory,'' 
Cambridge Univ. Press, 1998.

\bm{cvj}C.V. Johnson, ``D-Branes,'' 
Cambridge Univ. Press, 2003.

\bm{schwarz-review} 
 A.~Konechny and A.~Schwarz,
``Introduction to M(atrix) theory and noncommutative geometry,''
Phys.\ Rept.\  {\bf 360}, 353 (2002)
[arXiv:hep-th/0012145].

\bm{trieste}
N.~A.~Nekrasov,
``Trieste lectures on solitons in noncommutative gauge theories,''
arXiv:hep-th/0011095.

\bm{harvey}
J.~A.~Harvey,
``Komaba lectures on noncommutative solitons and D-branes,''
arXiv:hep-th/0102076.

\bm{nd}
M.~R.~Douglas and N.~A.~Nekrasov,
``Noncommutative field theory,''
Rev.\ Mod.\ Phys.\  {\bf 73}, 977 (2001)
[arXiv:hep-th/0106048].

\bm{szabo}
R.~J.~Szabo,
``Quantum field theory on noncommutative spaces,''
Phys.\ Rept.\  {\bf 378}, 207 (2003)
[arXiv:hep-th/0109162].

\bm{bfss}
T.~Banks, W.~Fischler, S.~H.~Shenker and L.~Susskind,
``M theory as a matrix model: A conjecture,''
Phys.\ Rev.\ D {\bf 55}, 5112 (1997)
[arXiv:hep-th/9610043].
 
\bm{taylor-M-review}
 W.~Taylor,
``M(atrix) theory: Matrix quantum mechanics as a fundamental theory,''
Rev.\ Mod.\ Phys.\  {\bf 73}, 419 (2001)
[arXiv:hep-th/0101126].
  
\bm{taylor-compact}
W.~I.~Taylor,
``D-brane field theory on compact spaces,''
Phys.\ Lett.\ B {\bf 394} (1997) 283
[arXiv:hep-th/9611042].

\bibitem{connes-rieffel}
A.~Connes and M.~A.~Rieffel,
``Yang-Mills For Noncommutative Two-Tori,''
Contemp.\ Math.\  {\bf 62} (1987) 237.

\bm{cheung}
Y.~K.~Cheung and M.~Krogh,
``Noncommutative geometry from 0-branes in a background B-field,''
Nucl.\ Phys.\ B {\bf 528} (1998) 185
[arXiv:hep-th/9803031].

\bm{douglas-hull}
M.~R.~Douglas and C.~M.~Hull,
``D-branes and the noncommutative torus,''
JHEP {\bf 9802}, 008 (1998)
[arXiv:hep-th/9711165].


\bm{schwarz-T}
A.~Schwarz,
``Morita equivalence and duality,''
Nucl.\ Phys.\ B {\bf 534} (1998) 720
[arXiv:hep-th/9805034].

\bm{zumino-T}
D.~Brace, B.~Morariu and B.~Zumino,
``Dualities of the matrix model from T-duality of the type II string,''
Nucl.\ Phys.\ B {\bf 545} (1999) 192
[arXiv:hep-th/9810099].

\bm{CH}
C.~S.~Chu and P.~M.~Ho,
``Noncommutative open string and D-brane,''
Nucl.\ Phys.\ B {\bf 550} (1999) 151
[arXiv:hep-th/9812219];
``Constrained quantization of open string in background B field and
Nucl.\ Phys.\ B {\bf 568} (2000) 447
[arXiv:hep-th/9906192].
\\
C.~S.~Chu,
``Noncommutative open string: Neutral and charged,''
arXiv:hep-th/0001144.

\bm{SW}
N.~Seiberg and E.~Witten,
``String theory and noncommutative geometry,''
JHEP {\bf 9909} (1999) 032
[arXiv:hep-th/9908142].

\bm{schomerus}
V.~Schomerus,
``D-branes and deformation quantization,''
JHEP {\bf 9906} (1999) 030
[arXiv:hep-th/9903205].


\bm{ars}
A.~Y.~Alekseev, A.~Recknagel and V.~Schomerus,
``Non-commutative world-volume geometries: Branes on SU(2) and fuzzy
spheres,''
JHEP {\bf 9909} (1999) 023
[arXiv:hep-th/9908040].

\bm{ncsym}
M.~M.~Sheikh-Jabbari,
``Super Yang-Mills theory on noncommutative torus from open strings
interactions,''
Phys.\ Lett.\ B {\bf 450} (1999) 119
[arXiv:hep-th/9810179].


\bm{1loop}
O.~Andreev and H.~Dorn,
``Diagrams of noncommutative Phi**3 theory from string theory,''
Nucl.\ Phys.\ B {\bf 583} (2000) 145
[arXiv:hep-th/0003113].\\
Y.~Kiem and S.~M.~Lee,
``UV/IR mixing in noncommutative field theory via open string loops,''
Nucl.\ Phys.\ B {\bf 586}, 303 (2000)
[arXiv:hep-th/0003145].\\
A.~Bilal, C.~S.~Chu and R.~Russo,
``String theory and noncommutative field theories at one loop,''
Nucl.\ Phys.\ B {\bf 582} (2000) 65
[arXiv:hep-th/0003180].\\
J.~Gomis, M.~Kleban, T.~Mehen, M.~Rangamani and S.~H.~Shenker,
``Noncommutative gauge dynamics from the string worldsheet,''
JHEP {\bf 0008}, 011 (2000)
[arXiv:hep-th/0003215]. \\
H.~Liu and J.~Michelson,
``Stretched strings in noncommutative field theory,''
Phys.\ Rev.\ D {\bf 62}, 066003 (2000)
[arXiv:hep-th/0004013].

\bm{dipole}
M.~M.~Sheikh-Jabbari,
``Open strings in a B-field background as electric dipoles,''
Phys.\ Lett.\ B {\bf 455} (1999) 129
[arXiv:hep-th/9901080].\\
D.~Bigatti and L.~Susskind,
``Magnetic fields, branes and noncommutative geometry,''
Phys.\ Rev.\ D {\bf 62} (2000) 066004
[arXiv:hep-th/9908056].

\bm{earlier}
M.~Li,
``Strings from IIB matrices,''
Nucl.\ Phys.\ B {\bf 499} (1997) 149
[arXiv:hep-th/9612222]. \\
D.~B.~Fairlie,
``Moyal brackets in M-theory,''
Mod.\ Phys.\ Lett.\ A {\bf 13} (1998) 263
[arXiv:hep-th/9707190].\\
T.~Curtright, D.~Fairlie and C.~K.~Zachos,
``Integrable symplectic trilinear interaction terms for matrix membranes,''
Phys.\ Lett.\ B {\bf 405} (1997) 37
[arXiv:hep-th/9704037].

\bm{reggeon}
C.~S.~Chu, R.~Russo and S.~Sciuto,
``Multiloop string amplitudes with B-field and noncommutative QFT,''
Nucl.\ Phys.\ B {\bf 585}, 193 (2000)
[arXiv:hep-th/0004183].

\bm{ncos1}
N.~Seiberg, L.~Susskind and N.~Toumbas,
``Strings in background electric field, space/time noncommutativity  and a new
noncritical string theory,''
JHEP {\bf 0006} (2000) 021
[arXiv:hep-th/0005040].

\bm{ncos2}
R.~Gopakumar, J.~M.~Maldacena, S.~Minwalla and A.~Strominger,
``S-duality and noncommutative gauge theory,''
JHEP {\bf 0006} (2000) 036
[arXiv:hep-th/0005048].

\bm{filk}
T.~Filk,
``Divergencies in a field theory on quantum space,''
Phys. Lett. B{\bf 376} (1996) 53.

\bm{iruv1}
S.~Minwalla, M.~Van Raamsdonk and N.~Seiberg,
``Noncommutative perturbative dynamics,''
JHEP {\bf 0002}, 020 (2000)
[arXiv:hep-th/9912072]. \\
M.~Van Raamsdonk and N.~Seiberg,
``Comments on noncommutative perturbative dynamics,''
JHEP {\bf 0003} (2000) 035
[arXiv:hep-th/0002186].

\bm{largeN}
H.~O.~Girotti, M.~Gomes, V.~O.~Rivelles and A.~J.~da Silva,
``The noncommutative supersymmetric nonlinear sigma model,''
Int.\ J.\ Mod.\ Phys.\ A {\bf 17} (2002) 1503
[arXiv:hep-th/0102101].
\\
E.~T.~Akhmedov, P.~DeBoer and G.~W.~Semenoff,
``Non-commutative Gross-Neveu model at large N,''
JHEP {\bf 0106} (2001) 009
[arXiv:hep-th/0103199].


\bm{wzw}
H.~O.~Girotti, M.~Gomes, V.~O.~Rivelles and A.~J.~da Silva,
``A consistent noncommutative field theory: The Wess-Zumino model,''
Nucl.\ Phys.\ B {\bf 587} (2000) 299
[arXiv:hep-th/0005272].

\bm{renorm}
I.~Chepelev and R.~Roiban,
``Renormalization of quantum field theories on noncommutative R**d.  I:
Scalars,''
JHEP {\bf 0005} (2000) 037
[arXiv:hep-th/9911098];
``Convergence theorem for non-commutative Feynman graphs and
renormalization,''
JHEP {\bf 0103} (2001) 001
[arXiv:hep-th/0008090].

\bibitem{u1}
J.~Gomis and T.~Mehen,
``Space-time noncommutative field theories and unitarity,''
Nucl.\ Phys.\ B {\bf 591} (2000) 265
[arXiv:hep-th/0005129].

\bm{u2}
L.~Alvarez-Gaume, J.~L.~F.~Barbon and R.~Zwicky,
``Remarks on time-space noncommutative field theories,''
JHEP {\bf 0105} (2001) 057
[arXiv:hep-th/0103069].

\bm{u3}
C.~S.~Chu, J.~Lukierski and W.~J.~Zakrzewski,
``Hermitian analyticity, IR/UV mixing and unitarity of noncommutative  field
theories,''
Nucl.\ Phys.\ B {\bf 632} (2002) 219
[arXiv:hep-th/0201144].

\bm{u4}
O.~Aharony, J.~Gomis and T.~Mehen,
``On theories with light-like noncommutativity,''
JHEP {\bf 0009} (2000) 023
[arXiv:hep-th/0006236].

\bm{wess}
J.~Madore, S.~Schraml, P.~Schupp and J.~Wess,
``Gauge theory on noncommutative spaces,''
Eur.\ Phys.\ J.\ C {\bf 16} (2000) 161
[arXiv:hep-th/0001203]. \\
B.~Jurco, S.~Schraml, P.~Schupp and J.~Wess,
``Enveloping algebra valued gauge transformations for non-Abelian gauge  groups
on non-commutative spaces,''
Eur.\ Phys.\ J.\ C {\bf 17} (2000) 521
[arXiv:hep-th/0006246].

\bm{sosp}
L.~Bonora, M.~Schnabl, M.~M.~Sheikh-Jabbari and A.~Tomasiello,
``Noncommutative SO(n) and Sp(n) gauge theories,''
Nucl.\ Phys.\ B {\bf 589} (2000) 461
[arXiv:hep-th/0006091]. \\
I.~Bars, M.~M.~Sheikh-Jabbari and M.~A.~Vasiliev,
``Noncommutative o*(N) and usp*(2N) algebras and the corresponding gauge  field
theories,''
Phys.\ Rev.\ D {\bf 64}, 086004 (2001)
[arXiv:hep-th/0103209].

\bm{wilson-line}
N.~Ishibashi, S.~Iso, H.~Kawai and Y.~Kitazawa,
``Wilson loops in noncommutative Yang-Mills,''
Nucl.\ Phys.\ B {\bf 573}, 573 (2000)
[arXiv:hep-th/9910004].

\bm{loopeqn}
M.~Abou-Zeid and H.~Dorn,
``Dynamics of Wilson observables in non-commutative gauge theory,''
Phys.\ Lett.\ B {\bf 504} (2001) 165
[arXiv:hep-th/0009231].
\\
H.~Dorn and A.~Torrielli,
``Loop equation in two-dimensional noncommutative Yang-Mills theory,''
JHEP {\bf 0401} (2004) 026
[arXiv:hep-th/0312047].

\bm{rey-w1}
S.~J.~Rey and R.~von Unge,
``S-duality, noncritical open string and noncommutative gauge theory,''
Phys.\ Lett.\ B {\bf 499} (2001) 215
[arXiv:hep-th/0007089]. 

\bm{gross-obs}
D.~J.~Gross, A.~Hashimoto and N.~Itzhaki,
``Observables of non-commutative gauge theories,''
Adv.\ Theor.\ Math.\ Phys.\  {\bf 4}, 893 (2000)
[arXiv:hep-th/0008075].


\bm{rey-w2}
S.~R.~Das and S.~J.~Rey,
``Open Wilson lines in noncommutative gauge theory and tomography of
holographic dual supergravity,''
Nucl.\ Phys.\ B {\bf 590} (2000) 453
[arXiv:hep-th/0008042].

\bibitem{DW}
A.~Dhar and S.~R.~Wadia, 
``A note on gauge invariant operators in noncommutative gauge theories  and the
matrix model,''
Phys.\ Lett.\ B {\bf 495} (2000) 413
[arXiv:hep-th/0008144].

\bm{sugra13}
H.~Liu and J.~Michelson,
``Ramond-Ramond couplings of noncommutative D-branes,''
Phys.\ Lett.\ B {\bf 518} (2001) 143
[arXiv:hep-th/0104139].\\
Y.~Okawa and H.~Ooguri,
``An exact solution to Seiberg-Witten equation of noncommutative gauge
theory,''
Phys.\ Rev.\ D {\bf 64} (2001) 046009
[arXiv:hep-th/0104036].\\
S.~Mukhi and N.~V.~Suryanarayana,
``Gauge-invariant couplings of noncommutative branes to Ramond-Ramond 
backgrounds,''
JHEP {\bf 0105} (2001) 023
[arXiv:hep-th/0104045].

\bm{liu-exact}
H.~Liu,
``*-Trek II: *n operations, open Wilson lines and the Seiberg-Witten  map,''
Nucl.\ Phys.\ B {\bf 614} (2001) 305
[arXiv:hep-th/0011125].

\bm{eff-s1}
Y.~Kiem, S.~J.~Rey, H.~T.~Sato and J.~T.~Yee,
``Open Wilson lines and generalized star product in nocommutative scalar
field theories,''
Phys.\ Rev.\ D {\bf 65} (2002) 026002
[arXiv:hep-th/0106121];
``Anatomy of one-loop effective action in noncommutative scalar field
theories,''
Eur.\ Phys.\ J.\ C {\bf 22} (2002) 757
[arXiv:hep-th/0107106].

\bm{eff-g1a}
M.~Van Raamsdonk,
``The meaning of infrared singularities in noncommutative gauge
theories,''
JHEP {\bf 0111} (2001) 006
[arXiv:hep-th/0110093].

\bm{eff-g1b}
A.~Armoni and E.~Lopez,
``UV/IR mixing via closed strings and tachyonic instabilities,''
Nucl.\ Phys.\ B {\bf 632} (2002) 240
[arXiv:hep-th/0110113].

\bm{eff-g1c}
Y.~J.~Kiem, Y.~J.~Kim, C.~Ryou and H.~T.~Sato,
``One-loop noncommutative U(1) gauge theory from bosonic worldline
approach,''
Nucl.\ Phys.\ B {\bf 630} (2002) 55
[arXiv:hep-th/0112176].
\\
J.~Levell and G.~Travaglini,
``Effective actions, Wilson lines and the IR/UV mixing in noncommutative
supersymmetric gauge theories,''
JHEP {\bf 0403} (2004) 021
[arXiv:hep-th/0308008].

\bm{eff-s2}
Y.~J.~Kiem, S.~S.~Kim, S.~J.~Rey and H.~T.~Sato,
``Anatomy of two-loop effective action in noncommutative field
theories,''
Nucl.\ Phys.\ B {\bf 641} (2002) 256
[arXiv:hep-th/0110066].

\bm{DBI12}
E.~S.~Fradkin and A.~A.~Tseytlin,
``Nonlinear Electrodynamics From Quantized Strings,''
Phys.\ Lett.\ B {\bf 163}, 123 (1985).\\
A.~Abouelsaood, C.~G.~.~Callan, C.~R.~Nappi and S.~A.~Yost,
``Open Strings In Background Gauge Fields,''
Nucl.\ Phys.\ B {\bf 280}, 599 (1987).\\
C.~G.~.~Callan, C.~Lovelace, C.~R.~Nappi and S.~A.~Yost,
``String Loop Corrections To Beta Functions,''
Nucl.\ Phys.\ B {\bf 288}, 525 (1987).

\bm{armoni}
A.~Armoni,
``Comments on perturbative dynamics of non-commutative Yang-Mills theory,''
Nucl.\ Phys.\ B {\bf 593}, 229 (2001)
[arXiv:hep-th/0005208].

\bm{renorm-gauge}
C.~P.~Martin and D.~Sanchez-Ruiz,
``The one-loop UV divergent structure of U(1) Yang-Mills theory on
noncommutative R**4,''
Phys.\ Rev.\ Lett.\  {\bf 83}, 476 (1999)
[arXiv:hep-th/9903077].\\
M.~M.~Sheikh-Jabbari,
``Renormalizability of the supersymmetric Yang-Mills theories on the
noncommutative torus,''
JHEP {\bf 9906} (1999) 015
[arXiv:hep-th/9903107].\\
T.~Krajewski and R.~Wulkenhaar,
``Perturbative quantum gauge fields on the noncommutative torus,''
Int.\ J.\ Mod.\ Phys.\ A {\bf 15} (2000) 1011
[arXiv:hep-th/9903187].

\bm{renorm-gaugeN}
L.~Bonora and M.~Salizzoni,
``Renormalization of noncommutative U(N) gauge theories,''
Phys.\ Lett.\ B {\bf 504} (2001) 80
[arXiv:hep-th/0011088].

\bm{iruv2}
M.~Hayakawa,
``Perturbative analysis on infrared and ultraviolet aspects of  noncommutative
QED on R**4,''
arXiv:hep-th/9912167.

\bm{CS}
C.~S.~Chu,
``Induced Chern-Simons and WZW action in noncommutative spacetime,''
Nucl.\ Phys.\ B {\bf 580} (2000) 352
[arXiv:hep-th/0003007].
\\
G.~H.~Chen and Y.~S.~Wu,
``One-loop shift in noncommutative Chern-Simons coupling,''
Nucl.\ Phys.\ B {\bf 593} (2001) 562
[arXiv:hep-th/0006114].

\bm{anom1}
F.~Ardalan and N.~Sadooghi,
``Axial anomaly in non-commutative QED on R**4,''
Int.\ J.\ Mod.\ Phys.\ A {\bf 16} (2001) 3151
[arXiv:hep-th/0002143].

\bibitem{anom2}
J.~M.~Gracia-Bondia and C.~P.~Martin,
``Chiral gauge anomalies on noncommutative R**4,''
Phys.\ Lett.\ B {\bf 479}, 321 (2000)
[arXiv:hep-th/0002171].

\bm{anom3}
L.~Bonora, M.~Schnabl and A.~Tomasiello,
``A note on consistent anomalies in noncommutative YM theories,''
Phys.\ Lett.\ B {\bf 485} (2000) 311
[arXiv:hep-th/0002210].

\bm{iruv3}
A.~Matusis, L.~Susskind and N.~Toumbas,
``The IR/UV connection in the non-commutative gauge theories,''
JHEP {\bf 0012}, 002 (2000)
[arXiv:hep-th/0002075].

\bm{gms}
R.~Gopakumar, S.~Minwalla and A.~Strominger,
``Noncommutative solitons,''
JHEP {\bf 0005} (2000) 020
[arXiv:hep-th/0003160].

\bm{soliton-d}
K.~Dasgupta, S.~Mukhi and G.~Rajesh,
``Noncommutative tachyons,''
JHEP {\bf 0006}, 022 (2000)
[arXiv:hep-th/0005006].
\\
J.~A.~Harvey, P.~Kraus, F.~Larsen and E.~J.~Martinec,
``D-branes and strings as non-commutative solitons,''
JHEP {\bf 0007} (2000) 042
[arXiv:hep-th/0005031].
\\
E.~Witten,
``Noncommutative tachyons and string field theory,''
arXiv:hep-th/0006071.
\\
M.~Aganagic, R.~Gopakumar, S.~Minwalla and A.~Strominger,
``Unstable solitons in noncommutative gauge theory,''
JHEP {\bf 0104} (2001) 001
[arXiv:hep-th/0009142].
\\
J.~A.~Harvey, P.~Kraus and F.~Larsen,
``Exact noncommutative solitons,''
JHEP {\bf 0012} (2000) 024
[arXiv:hep-th/0010060].

\bm{ncsol-exact}
A.~P.~Polychronakos,
``Flux tube solutions in noncommutative gauge theories,''
Phys.\ Lett.\ B {\bf 495} (2000) 407
[arXiv:hep-th/0007043].
\\
D.~J.~Gross and N.~A.~Nekrasov,
``Dynamics of strings in noncommutative gauge theory,''
JHEP {\bf 0010} (2000) 021
[arXiv:hep-th/0007204].
\\
D.~Bak,
``Exact multi-vortex solutions in noncommutative Abelian-Higgs theory,''
Phys.\ Lett.\ B {\bf 495} (2000) 251
[arXiv:hep-th/0008204].

\bibitem{TZ-rev}
W.~Taylor and B.~Zwiebach,
``D-branes, tachyons, and string field theory,''
arXiv:hep-th/0311017.
\\
 A.~Sen,
``Tachyon dynamics in open string theory,''
arXiv:hep-th/0410103.

\bm{ns-ins}
N.~Nekrasov and A.~Schwarz,
``Instantons on noncommutative R**4 and (2,0) superconformal six  dimensional
theory,''
Commun.\ Math.\ Phys.\  {\bf 198}, 689 (1998)
[arXiv:hep-th/9802068].
 
\bm{mono}
D.~J.~Gross and N.~A.~Nekrasov,
``Monopoles and strings in noncommutative gauge theory,''
JHEP {\bf 0007} (2000) 034
[arXiv:hep-th/0005204].


\bm{kappa}
J.~Lukierski, H.~Ruegg, A.~Nowicki and V.~N.~Tolstoi,
``Q deformation of Poincare algebra,''
Phys.\ Lett.\ B {\bf 264} (1991) 331;
``New quantum Poincare algebra and k deformed field theory,''
Phys.\ Lett.\ B {\bf 293} (1992) 344.

\bm{QG}
see for example, M.~Chaichian and A.~P.~Demichev,
``Introduction to quantum groups,'' Singapore, World Scientific, 1996.

\bm{gp13}
J.~de Boer, P.~A.~Grassi and P.~van Nieuwenhuizen,
``Non-commutative superspace from string theory,''
Phys.\ Lett.\ B {\bf 574} (2003) 98
[arXiv:hep-th/0302078].
\\
H.~Ooguri and C.~Vafa,
``The C-deformation of gluino and non-planar diagrams,''
Adv.\ Theor.\ Math.\ Phys.\  {\bf 7} (2003) 53
[arXiv:hep-th/0302109];
``Gravity induced C-deformation,''
Adv.\ Theor.\ Math.\ Phys.\  {\bf 7} (2004) 405
[arXiv:hep-th/0303063].
\\
N.~Berkovits and N.~Seiberg,
``Superstrings in graviphoton background and N = 1/2 + 3/2 supersymmetry,''
JHEP {\bf 0307} (2003) 010
[arXiv:hep-th/0306226].
\\
M.~Billo, M.~Frau, I.~Pesando and A.~Lerda,
``N = 1/2 gauge theory and its instanton moduli space from open strings in R-R
background,''
JHEP {\bf 0405} (2004) 023
[arXiv:hep-th/0402160].
\\
M.~Billo, M.~Frau, F.~Lonegro and A.~Lerda,
``N = 1/2 quiver gauge theories from open strings with R-R fluxes,''
arXiv:hep-th/0502084.

\bm{seiberg-ncsusy}
N.~Seiberg,
``Noncommutative superspace, N = 1/2 supersymmetry, field theory and  string
theory,''
JHEP {\bf 0306} (2003) 010
[arXiv:hep-th/0305248].

\bm{ncsusy-wz}
R.~Britto, B.~Feng and S.~J.~Rey,
``Deformed superspace, N = 1/2 supersymmetry and (non)renormalization
theorems,''
JHEP {\bf 0307} (2003) 067
[arXiv:hep-th/0306215].
\\
M.~T.~Grisaru, S.~Penati and A.~Romagnoni,
``Two-loop renormalization for nonanticommutative N = 1/2 supersymmetric  WZ
model,''
JHEP {\bf 0308} (2003) 003
[arXiv:hep-th/0307099].
\\
R.~Britto and B.~Feng,
``N = 1/2 Wess-Zumino model is renormalizable,''
Phys.\ Rev.\ Lett.\  {\bf 91} (2003) 201601
[arXiv:hep-th/0307165].
\\
A.~Romagnoni,
``Renormalizability of N = 1/2 Wess-Zumino model in superspace,''
JHEP {\bf 0310} (2003) 016
[arXiv:hep-th/0307209].

\bm{ncsusy-g}
O.~Lunin and S.~J.~Rey,
``Renormalizability of non(anti)commutative gauge theories with N = 1/2
supersymmetry,''
JHEP {\bf 0309} (2003) 045
[arXiv:hep-th/0307275].
\\
D.~Berenstein and S.~J.~Rey,
``Wilsonian proof for renormalizability of N = 1/2 supersymmetric field
theories,''
Phys.\ Rev.\ D {\bf 68} (2003) 121701
[arXiv:hep-th/0308049].

\bm{int-field}
O.~Lechtenfeld, A.~D.~Popov and B.~Spendig,
``Noncommutative solitons in open N = 2 string theory,''
JHEP {\bf 0106} (2001) 011
[arXiv:hep-th/0103196].\\
O.~Lechtenfeld and A.~D.~Popov,
``Noncommutative multi-solitons in 2+1 dimensions,''
JHEP {\bf 0111} (2001) 040
[arXiv:hep-th/0106213].\\
 M.T.Grisaru and S.Penati,
``The noncommutative sine-Gordon system,''
Nucl. Phys. B 655 (2003) 250 [arXiv:hep-th/0112246].\\
 M.Wolf,
``Soliton antisoliton scattering configurations in a noncommutative sigma
model in 2+1 dimensions,''JHEP 0206 (2002) 055 [arXiv:hep-th/0204185].\\
I.~Cabrera-Carnero and M.~Moriconi,
``Noncommutative integrable field theories in 2d,''
Nucl.\ Phys.\ B {\bf 673} (2003) 437
[arXiv:hep-th/0211193].\\
M.Ihl and S.Uhlmann,
``Noncommutative extended waves and soliton-like configurations in N=2
string theory,'' Int.J.Mod.Phys. A 18 (2003) 4889 [arXiv:hep-th/0211263].\\
O.~Lechtenfeld, L.~Mazzanti, S.~Penati, A.~D.~Popov and L.~Tamassia,
``Integrable noncommutative sine-Gordon model,''
Nucl.\ Phys.\ B {\bf 705} (2005) 477
[arXiv:hep-th/0406065].

\bm{ward}
R.~S.~Ward,
``Integrable And Solvable Systems, And Relations Among Them,''
Phil.\ Trans.\ Roy.\ Soc.\ Lond.\ A {\bf 315} (1985) 451.

\bm{lax}
M.~Hamanaka and K.~Toda,
``Towards noncommutative integrable systems,''
Phys.\ Lett.\ A {\bf 316} (2003) 77
[arXiv:hep-th/0211148].
\\
M.~Hamanaka,
``Commuting flows and conservation laws for noncommutative Lax  hierarchies,''
arXiv:hep-th/0311206.\\
A.~Dimakis and F.~Mueller-Hoissen,
``Extension of noncommutative soliton hierarchies,''
J.\ Phys.\ A {\bf 37} (2004) 4069
[arXiv:hep-th/0401142].

\bm{phen}
The literature is vast. See for example,
I.~Mocioiu, M.~Pospelov and R.~Roiban,
``Low-energy limits on the antisymmetric tensor field background on the
brane and on the non-commutative scale,''
Phys.\ Lett.\ B {\bf 489} (2000) 390
[arXiv:hep-ph/0005191].
\\
J.~L.~Hewett, F.~J.~Petriello and T.~G.~Rizzo,
``Signals for non-commutative interactions at linear colliders,''
Phys.\ Rev.\ D {\bf 64}, 075012 (2001)
[arXiv:hep-ph/0010354].
\\
V.~V.~Khoze and G.~Travaglini,
``Wilsonian effective actions and the IR/UV mixing in noncommutative  gauge
theories,''
JHEP {\bf 0101} (2001) 026
[arXiv:hep-th/0011218].
\\
C.~S.~Chu, V.~V.~Khoze and G.~Travaglini,
``Dynamical breaking of supersymmetry in noncommutative gauge theories,''
Phys.\ Lett.\ B {\bf 513} (2001) 200
[arXiv:hep-th/0105187].
\\
Z.~Guralnik, R.~Jackiw, S.~Y.~Pi and A.~P.~Polychronakos,
``Testing non-commutative QED, constructing non-commutative MHD,''
Phys.\ Lett.\ B {\bf 517} (2001) 450
[arXiv:hep-th/0106044].
\\
A.~Anisimov, T.~Banks, M.~Dine and M.~Graesser,
``Comments on non-commutative phenomenology,''
Phys.\ Rev.\ D {\bf 65}, 085032 (2002)
[arXiv:hep-ph/0106356].
\\
C.~E.~Carlson, C.~D.~Carone and R.~F.~Lebed,
``Bounding noncommutative QCD,''
Phys.\ Lett.\ B {\bf 518}, 201 (2001)
[arXiv:hep-ph/0107291].
\\
X.~Calmet, B.~Jurco, P.~Schupp, J.~Wess and M.~Wohlgenannt,
``The standard model on non-commutative space-time,''
Eur.\ Phys.\ J.\ C {\bf 23} (2002) 363
[arXiv:hep-ph/0111115].

\bm{cosmo}
See for example,
C.~S.~Chu, B.~R.~Greene and G.~Shiu,
``Remarks on inflation and noncommutative geometry,''
Mod.\ Phys.\ Lett.\ A {\bf 16} (2001) 2231
[arXiv:hep-th/0011241].
\\
F.~Lizzi, G.~Mangano, G.~Miele and M.~Peloso,
``Cosmological perturbations and short distance physics from
noncommutative geometry,''
JHEP {\bf 0206} (2002) 049
[arXiv:hep-th/0203099].
\\
R.~Brandenberger and P.~M.~Ho,
``Noncommutative spacetime, stringy spacetime uncertainty principle, and
density fluctuations,''
Phys.\ Rev.\ D {\bf 66} (2002) 023517
[AAPPS Bull.\  {\bf 12N1} (2002) 10]
[arXiv:hep-th/0203119].
\\
S.~F.~Hassan and M.~S.~Sloth,
``Trans-Planckian effects in inflationary cosmology and the modified
uncertainty principle,''
Nucl.\ Phys.\ B {\bf 674} (2003) 434
[arXiv:hep-th/0204110].


\bm{adscft}
See for example the reviews,
O.~Aharony, S.~S.~Gubser, J.~M.~Maldacena, H.~Ooguri and Y.~Oz,
``Large N field theories, string theory and gravity,''
Phys.\ Rept.\  {\bf 323} (2000) 183
[arXiv:hep-th/9905111].\\
E.~D'Hoker and D.~Z.~Freedman,
``Supersymmetric gauge theories and the AdS/CFT correspondence,''
arXiv:hep-th/0201253.

\bm{ncadscft} 
A.~Hashimoto and N.~Itzhaki,
``Non-commutative Yang-Mills and the AdS/CFT correspondence,''
Phys.\ Lett.\ B {\bf 465} (1999) 142
[arXiv:hep-th/9907166]. \\
J.~M.~Maldacena and J.~G.~Russo,
``Large N limit of non-commutative gauge theories,''
JHEP {\bf 9909} (1999) 025
[arXiv:hep-th/9908134].

\bm{ncadscft2}
M.~Li and Y.~S.~Wu,
``Holography and noncommutative Yang-Mills,''
Phys.\ Rev.\ Lett.\  {\bf 84} (2000) 2084
[arXiv:hep-th/9909085].
\\
M.~Alishahiha, Y.~Oz and M.~M.~Sheikh-Jabbari,
``Supergravity and large N noncommutative field theories,''
JHEP {\bf 9911} (1999) 007
[arXiv:hep-th/9909215].
\\
R.~G.~Cai and N.~Ohta,
``On the thermodynamics of large N non-commutative super Yang-Mills  theory,''
Phys.\ Rev.\ D {\bf 61} (2000) 124012
[arXiv:hep-th/9910092];
``Noncommutative and ordinary super Yang-Mills on (D(p-2),Dp) bound  states,''
JHEP {\bf 0003} (2000) 009
[arXiv:hep-th/0001213];
``(F1, D1, D3) bound state, its scaling limits and SL(2,Z) duality,''
Prog.\ Theor.\ Phys.\  {\bf 104} (2000) 1073
[arXiv:hep-th/0007106].
\\
S.~R.~Das, S.~Kalyana Rama and S.~P.~Trivedi,
``Supergravity with self-dual B fields and instantons in noncommutative  gauge
theory,''
JHEP {\bf 0003} (2000) 004
[arXiv:hep-th/9911137].


\end{thebibliography}
\end{document}